\documentclass[a4paper]{article}
\pdfoutput=1
\usepackage{multirow}
\usepackage{booktabs}
\usepackage{INTERSPEECH2020}
\usepackage{verbatimbox}
\usepackage{setspace}

\usepackage{color}
\title{NPU Speaker Verification System for INTERSPEECH 2020 Far-Field Speaker Verification Challenge}
\name{Li Zhang, Jian Wu, Lei Xie{*}  \thanks{* Corresponding author.} }
\address{
  Audio, Speech and Language Processing Group (ASLP@NPU), School of Computer Science, Northwestern Polytechnical University, Xi'an, China
  }
\email{lizhang.aslp.npu@gmail.com, lxie@nwpu.edu.cn}
\begin{document}
\maketitle

\setlength{\baselineskip}{10.2pt}
\begin{abstract}
  This paper describes the NPU system submitted to Interspeech 2020 Far-Field Speaker Verification Challenge (FFSVC). We particularly focus on far-field text-dependent SV from single (task1) and multiple microphone arrays (task3). The major challenges in such scenarios are \textit{short utterance} and \textit{cross-channel and distance mismatch} for enrollment and test. With the belief that better speaker embedding can alleviate the effects from short utterance, we introduce a new speaker embedding architecture - ResNet-BAM, which integrates a bottleneck attention module with ResNet as a simple and efficient way to further improve the representation power 
 of ResNet. This contribution brings up to 1\% EER reduction. We further address the mismatch problem in three directions. First, \textit{domain adversarial training}, which aims to learn domain-invariant features, can yield to 0.8\% EER reduction. Second, \textit{front-end signal processing}, including WPE and beamforming, has no obvious contribution, but together with data selection and domain adversarial training, can further contribute to 0.5\% EER reduction. Finally, data augmentation, which works with a specifically-designed data selection strategy, can lead to 2\% EER reduction. Together with the above contributions, in the middle challenge results, our single submission system (without multi-system fusion) achieves the first and second place on task 1 and task 3, respectively. 
\end{abstract}
 
\section{Introduction}
With the rise of deep neural networks (DNN) and easy availability of computing resources and massive data, speaker verification (SV) performance has been significantly improved in the past several years. However, such advances are mainly achieved on close-talk scenarios with less interference. With the fast proliferating of smart devices, such as smart speakers and various voice-enabled IoT gadgets, the need for far-field speech interaction will continue to grow. Recognizing who is speaking is essential to such smart devices to provide customized services. Far-field speech tasks including speaker recognition remain challenging yet due to attenuated speech signals, noise interference as well as room reverberations. 
Particularly for smart devices, it is more convenient for users to enroll on a short utterance, e.g., a trigger word, from a close-talk portable device such as a cellphone but talk to a smart device from distance to obtain authentication. It apparently raises other problems -- short utterance verification, data mismatch between enroll and test in terms of channels and distances. Interspeech 2020 Far-Field Speaker Verification Challenge (FFSVC)~\cite{qin2020ffsvc} provides a common testbed for researchers to address the above mentioned difficult problems -- \textit{deteriorated signal}, \textit{short utterance} and \textit{data mismatch}. 

In this paper, we present our efforts to deal with the above mentioned problems with our submitted system to FFSVC. We particularly introduce our approaches in the two text-dependent tasks, i.e., far-field text-dependent SV from single (task1) and multiple arrays (task3). We introduce a new speaker embedding architecture with more powerful speaker representation, which is built on ResNet with attention module. The new architecture ResNet-BAM achieves 1\% abosolute equal error rate (EER) reduction compared to the baseline ResNet model. We further address deteriorated signal and mismatch problem with front-end processing and domain adversarial training (DAT). The two methods can bring 0.5\% and 0.8\% absolute EER reduction, respectively. Finally, data augmentation, which works with a specifically-designed data selection strategy, can lead to 2\% absolute EER reduction. With the above contributions, our single submission system (without multi-system fusion) achieves the first and second place on task 1 and task 3, respectively.

The rest is organized as follows. Section 2 introduces the related works and Section 3 describes the system overview. Section 4 details the proposed Resnet-BAM model for better speaker embedding, followed by experiments to validate its efficacy. Section 5 focuses on domain adversarial training and its evaluation. Section 6 analyzes the effects from front-end processing and data augmentation and selection. Section 7 summarizes the official evaluation results and concludes this paper. 
\setlength{\intextsep}{4.5pt}

\setlength{\textfloatsep}{4.0pt}
\setlength{\abovecaptionskip}{1.5pt}
\setlength{\belowcaptionskip}{0.7pt}
\section{Related Works}

Most of approaches to deal with short utterance SV task focus on improving the speaker embedding network with stronger extracting capabilities. Some improve x-vector-based models~\cite{kanagasundaram2019study} while others work on start-of-the-art convolution neural networks (CNN) on various datasets~\cite{yadav2019frequency,zhou2019cnn}. Li \textit{et al.}~\cite{li2017deep} have reported that CNN-based network even can recognize speaker by cough or laugh recordings~\cite{zhang2017speaker}, which are extremely short `utternaces'. Wang \textit{et al.}~\cite{wang2019discriminative} integrated deep discriminant analysis into CNN-based structure to achieve good performance on short utterance SV. Study from~\cite{gusev2020deep} has confirmed that ResNet architectures outperform the standard x-vector approach in terms of SV quality for both long-duration and short-duration utterances. Since then, ResNet has become the most popular network structure for speaker embedding extraction. As loss function is also essential to the network's learning ability, there have been various studies exploring in this direction~\cite{coria2020comparison}.

Training-testing or enroll-testing mismatch is a research problem explored for many years. Model adaptation or more formally domain adaptation, which aims to transfer the source model to the target domain, has been studied extensively~\cite{alam2018speaker,tu2019variational}. Domain adversarial training (DAT) is the most recent approaches developed with DNN's strong and flexible modeling ability. As a specifically designed multi-task learning framework, domain adversarial neural network (DANN) injects a domain classifier as an auxiliary task with a gradient reversal layer to learn domain-invariant features~\cite{sun2017unsupervised}. Wang \textit{et al.}~\cite{wang2018unsupervised} have recently applied DANN to remove cross-dataset variation and project data from difference datasets into the same subspace and superior SV performance has been reported on 2013 domain adaptation challenge (DAC) data. Many followers along this direction have investigated on multi-language ~\cite{rohdin2019speaker,xia2019cross} and multi-channel adaptation~\cite{shon2017autoencoder,luu2020channel}.

The mismatch from cross-channel (and distance) enrollment and test usually can be compensated by so-called \textit{front-end processing} which utilizes traditional signal processing technologies. A dereverberation module, such as the one adopts weighted prediction error (WPE)~\cite{cai2019dku} algorithm, is usually employed to remove reverberations from far-field collected speech signal thus to match the cross-talk signal. Moreover, beamformers~\cite{taherian2019deep} are adopted to process multi-channel signals collected from microphone array(s), resulting in a single-channel enhanced speech signal. There are also some new development on neural front-ends, such as neural dereverberation \cite{kinoshita2017neural} and neural beamforming~\cite{li2016neural,drude2018integrating}. Yang \textit{et al.}~\cite{yang2019joint} have proposed joint optimization of neural beamforming and dereverberation with x-vectors for robust speaker verification.

Data augmentation~\cite{qin2020svc} is another commonly used simple-but-effective trick to alleviate data mismatch. For instance, many approaches on VOiCEs challenge~\cite{cai2019dku,novoselov2019stc,burget2019analysis} have considered this trick with improved performances. Augmentation to the training data makes the model `see' more acoustic environments with diversity, leading to more robust speaker embedding~\cite{snyder2018x}. Meanwhile, augmentation in enrollment and test not only can compensate the mismatch between enrollment and test, but also can make up for the negative effects from short utterance duration during evaluation. So far, data augmentation has been an effective and intuitive way for robust modeling by improving the diversity of the data. But there is still a fundamental problem: does all the augmented data work well and how to select more effective data? This paper  tries to answer the question with a simple data selection strategy.

\section{System Overview}
\begin{figure}[h]
  \centering
  \includegraphics [width=\linewidth]{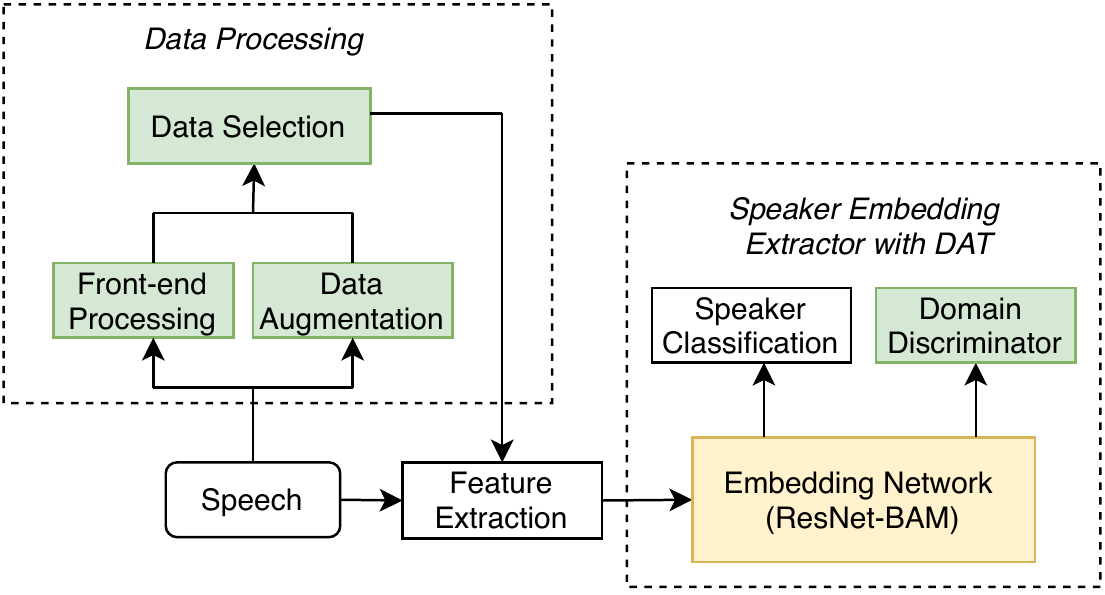}
  \caption{The overview of our speaker verification system.}
  \label{fig:pain_overview}
\end{figure}

Figure \ref{fig:pain_overview} illustrates the basic diagram of our system in FFSVC. It mainly consists of two sub-modules. Aiming to alleviate the data mismatch problem, the data processing module is composed of  front-end processing and data augmentation, while both go through data selection to result in the final `high-quality' augmented data. Another core module is speaker embedding extractor, which is composed of the proposed ResNet-BAM network for extracting better speaker embedding and the deep adversarial training built upon the embedding network to learn domain-invariant and speaker-discriminative features.

\section{ResNet-BAM Model}
\begin{figure}[h]
  \centering
  \includegraphics [width=\linewidth]{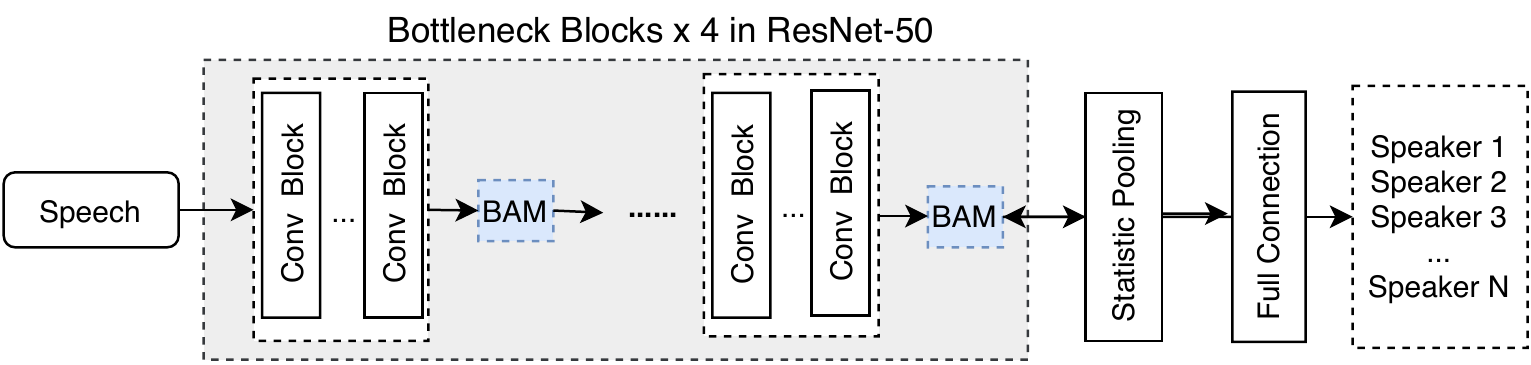}
  \caption{The structrue of ResNet-BAM}
  \label{fig:resnet_bam}
\end{figure}

Our speaker embedding extractor is a CNN-based ResNet-50 model. We replace the average pooling layer of ResNet-50 \cite{he2016deep} with a statistic pooling layer the same as that in x-vector~\cite{snyder2018x}, and then add a fully connected layer followed with the statistic pooling layer as our baseline model.  We propose to use an attention-enhanced ResNet structure to further improve the ability of the baseline embedding extractor, which originally shows superior performance in several image classification and detection tasks~\cite{park2018bam}. Specifically, we add bottleneck attention modules (BAM)~\cite{park2018bam} followed with bottleneck layers (ResNet-BAM) in ResNet-50 to extract better speaker embedding. BAM is able to emphasize important elements in 3D feature map generated from convolution. In speech, 3D feature map has channel dimension (filter number of convolution), time dimension and frequency dimension. There are two branches to calculate attention masks: channel attention is to learn which channels are more important for the final classification task, while time-frequency attention aims to learn which points in time-frequency domain are more effective for the classification task. The two branches (channels and time-frequency) explicitly learn `what' and `where' in the spectral graph to focus on. The structure of ResNet-BAM is shown in Figure~\ref{fig:resnet_bam}.

\subsection{Attention module}
\begin{figure}[h]
  \centering
  \includegraphics [width=\linewidth]{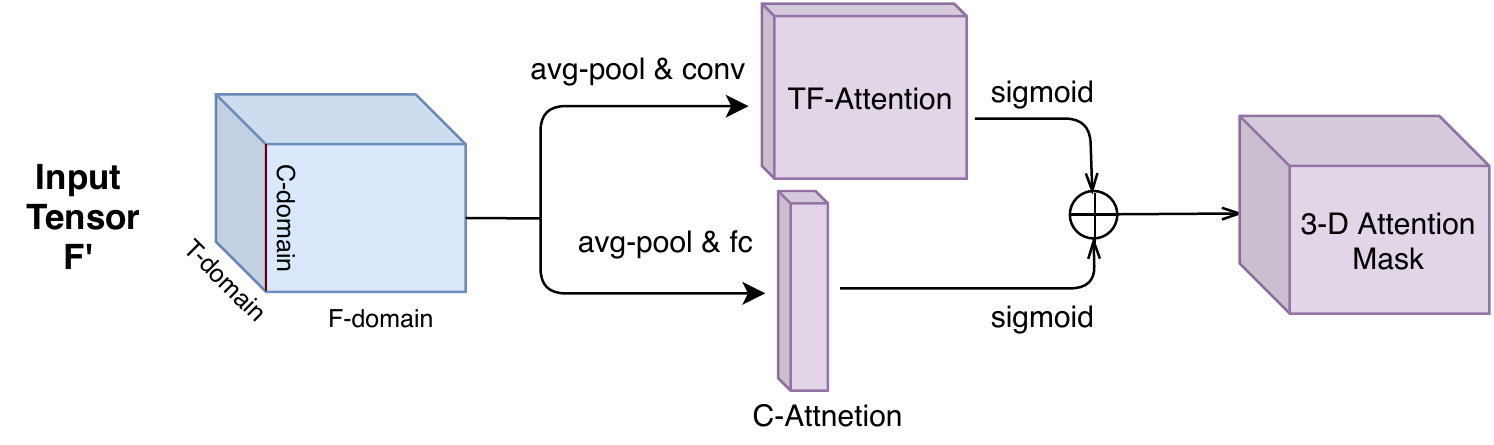}
  \caption{Details of bottleneck attention module (BAM) in ResNet-BAM. $F'$ is  bottleneck output of several convolution.}
  \label{fig:attention}
\end{figure}

The detail of bottleneck attention module in ResNet-BAM is illustrated in Figure~\ref{fig:attention}. 
After several layers of convolution on input $x$, we obtain the 3D feature map $F' \in  R^{C \times T \times F}$ after the bottleneck layer. Then an attention module infers a 3D attention map $ M(F)  \in  R^{H\times T\times F} $. The refined feature map after the attention module $F''$ is computed as
\begin{footnotesize}
\begin{equation}
F'' = F'+F'\cdot M(F')
\end{equation}
\end{footnotesize}where $\cdot$ denotes element-wise multiplication. $M(F')$ is combined with two attentions masks -- channel attention mask  $M_{C}(F')\in R^{C}$ and time-frequency attention mask $ M_{tf}(F')\in R^{T\times F}$. Two branches of attention are computed in parallel. In the original ResNet-BAM~\cite{park2018bam}, $M(F')$ is the direct addition between $M_{tf}(F')$ and $M_{tf}(F')$, and then normalized by sigmoid function into a 3D attention mask in range of (0,1). But according to our empirical experiments, there exists an offset problem if corresponding elements of the two attention masks have different sign, and we cannot judge the direct relationship between the positive or negative values of the attention masks and the final recognition result. Hence we switch the two steps: first do sigmoid on the two masks and then add together. Finally the attention mask $M(F')$ is calculated as
 \begin{footnotesize}
 \begin{equation}
 M(F')=(\text{Sigmoid}(M_{C}(F')) + \text{Sigmoid}(M_{tf}(F'))/2.
\end{equation}
\end{footnotesize}Although each channel (filters in convolution) contains a specific feature representation, different channel elements cannot have the same effect on the final recognition task. Learning the importance mask of each channel elements thus not only can guide the model to focus on the more effective points but also speed up model convergence. Given a 3D feature map $F'$, we use global average pooling to get a vector in channel dimension.  Then multi-layer perceptron (MLP) is utilized to estimate attention across channels and after batch normalization, the output is produced. In brief, the channel attention is computed as
\begin{footnotesize}
\begin{equation}
M_{C}(F') = \text{BN}(\text{MLP}(\text{AvgPool}(F')_{C \times 1 \times 1})) 
\end{equation}
\end{footnotesize}The time-frequency (TF) branch aims to learn an attention map to emphasize or suppress different points on the spectral graph. $M_{tf}(F')$ is calculated in time-frequency domain. Firstly, average pooling is conducted on channel dimension, then we use convolutions to learn a  2D mask in TF domain. After we normalize the time-frequency mask. The TF mask is calculated as
\begin{footnotesize}

\begin{equation}
M_{tf}(F') = \text{BN}(f'(\text{AvgPool}(F')_{1 \times T \times F}))
\end{equation}
\end{footnotesize}where $f'$ represents the convolution operations after average pooling on 3D input feature.

Finally, channel-attention and TF-attention masks are combined through Eq.(2). 

\subsection{Experiments on ResNet-BAM}
The challenge only allows to use the datasets shared on OpenSLR for model training. Our team choose five datasets (SLR 33, 38, 62, 82 and 85) together with the FFSVC official data (FFSVC20) as the basic training data, in which total speaker number is 3,211 with about 2,100 hours of Mandarin speech. The official development set includes 35 speakers. Trials are 53,996 pairs in both task 1 and 3. Enrollment data is recorded from iPhone. Test data in task 1 is from one random selected microphone array with four channels while test data in task 3 is from 2-4 random selected microphone arrays, each array with four channels~\cite{qin2020ffsvc}.

We conduct all experiments on Pytorch. Acoustic feature is 30-dim MFCCs with kaldi~\cite{Povey_ASRU2011} energy VAD to remove silence frames beforehand. Batch size is set to 64 and input tensor size is [1,256,30]. We trunk every utterance randomly. If the audio not reaches to 256 frames, we repeat the original audio and random trunk again. Initial learning rate is 0.1 and it decays to the original 10\% every 5 epochs. Optimizer is Stochastic gradient descent in Pytorch. The core metrics of the challenge are equal error rate (EER) and minimum detection cost function (minDCF). All scores of this paper we calculated are based on cosine distance.

Results on development data for task 1 and 3 are summarized in Table~\ref{tab:baseline-results}. We can see that the proposed ResNet-BAM model can bring roughly 1\% reduction in EER for both task 1 and task 3. We believe that the performance gain mainly comes from the improved speaker representation power by using the specifically-designed attention module which can guide the model to learn discriminative embedding more effectively for the speaker verification task.

\begin{table}[]
\setlength{\abovecaptionskip}{0.2cm}
\scriptsize
\caption{ Results on development set for two embedding networks (ResNet and ResNet-BAM) w/o and w/ DAT.
}
\label{tab:baseline-results}
\begin{tabular}{lllll}
\hline
\multicolumn{1}{l|}{\multirow{2}{*}{Model Name}} &  \multicolumn{2}{l|}{Task1} & \multicolumn{2}{l}{Task3} \\ \cline{2-5} 
\multicolumn{1}{l|}{} & EER (\%) & \multicolumn{1}{l|}{minDCF} & EER (\%) & minDCF \\ \hline
ResNet & 8.34 & 0.8539 & 7.98 & 0.8231 \\
ResNet-BAM & \textbf{7.43} & \textbf{0.7707 }& \textbf{6.89} & \textbf{0.7312} \\ \hline
ResNet-DAT & 7.59 & 0.7921 & 7.05 & 0.7230 \\
ResNet-BAM-DAT & \textbf{6.71} & \textbf{0.7507} & \textbf{6.19} & \textbf{0.7023} \\ \hline
\end{tabular}
\end{table}

\section{Domain Adversarial Training}
To alleviate domain mismatch, an intuitive idea is to project two different domains into a common space for speaker recognition. This can be achieved by domain adversarial training (DAT) with a gradient reversal layer (GRL) which aims to learn domain-invariant and discriminative speaker embedding~\cite{wang2018unsupervised}. 

\subsection{Invariant feature learning via DAT}
\begin{figure}[h]
   \centering
   \includegraphics[width=\linewidth]{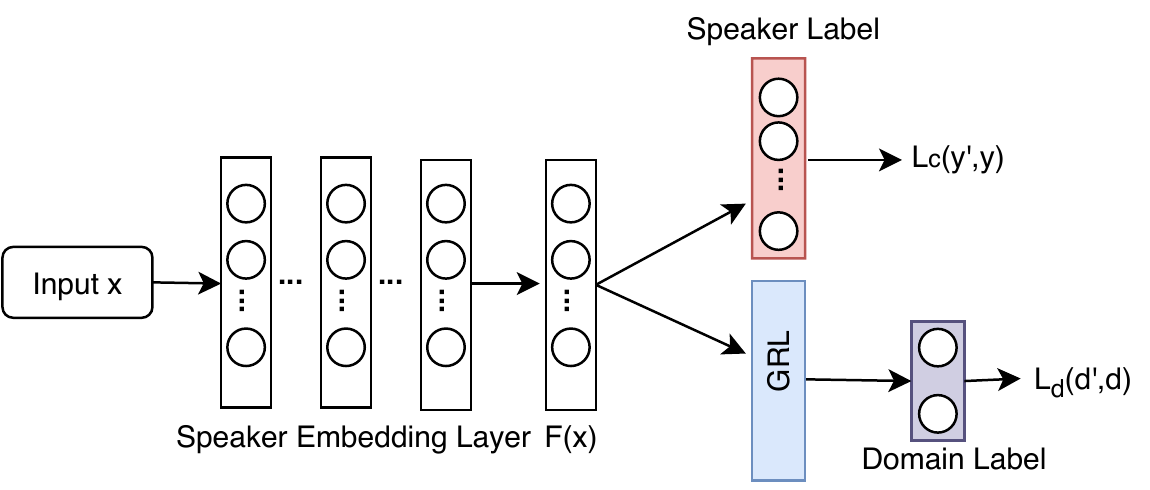}
   \caption{DAT in speaker embedding network. GRL is a gradient reverse layer.}
   \label{fig:dat}

\end{figure}

As shown in Figure~\ref{fig:dat}, the DAT module is built on a pre-trained ResNet-BAM model, formed as a multi-task learning (MTL) problem, where the main task is speaker recognition and the auxiliary task is domain discrimination. Specifically, the domain discriminator sub-network in our task is designed to distinguish close-talk speech from far-field speech. The two branches take input from a shared feature extractor sub-network that aims to learn representations that capture the underlying speaker discriminative information and are independent of speech domain. Different from conventional MTL, an inserted gradient reversal layer is essential to learn domain-independent features.

 Given an input ${x}$ and its speaker label ${y}$ as well domain label $d$, the predicted speaker and domain label are ${y'}$ and ${d'}$, respectively. The loss of the two tasks are combined as
 \begin{footnotesize}
 \begin{equation}
L(\theta_{f},\theta_{y},\theta_{d}) = \frac{1}{n}\sum_{i=1}^{n}L_{y}^{i}(\theta_{f},\theta_{y})-\lambda \frac{1}{n}\sum_{i=1}^{n}L_{y}^{i}(\theta_{f},\theta_{d}) 
\end{equation}
\end{footnotesize} 
where $\theta_{f},\theta_{y},\theta_{d}$ are parameters of the shared feature extractor and the two classifiers, and $L_{y}$ and $L_{d}$ are the speaker prediction loss and the domain classification loss, respectively. $n$ is the number of training samples. The joint loss is to minimize the speaker classification loss and maximize the domain discriminator loss at the same time, which is achieved by the GRL to reverse the sign of gradient before the domain discriminator. By this way, the feature extractor is able to learn domain-invariant and speaker-discriminative features.

\subsection{Experiments on DAT} 
Our DAT approach is based on the pre-trained ReseNet and ReseNet-BAM models in Section 4. Recall that the two models are trained using the datasets introduced in Section 4.2. Then the two models are equipped with the DAT structure and fine-tuned using the official FFSVC20 data. Here, data recorded on iPhone is regard as source while data recorded by microphone array(s) is considered as target. Results in Table~\ref{tab:baseline-results} shows that with the help of DAT, EER for both tasks has been reduced  about 0.8\%. This performance gain can be observed for both ReseNet and ReseNet-BAM models. Using this DAT approach, we are able to minimize the gap between the source and target feature distributions. Therefore, the learned embedding is less dependent on the domain shift.


\section{Front-end and Data Augmentation}

\subsection{Front-end processing}

In evaluation stage, we use the WPE algorithm~\cite{nakatani2010speech} to handle
reverberation issues. As the test data provide multi-channel 
speech signal, we adopt minimum variance distortionless response (MVDR) beamformer to suppress
interfering noise, which has been proved to benifit other speech tasks, e.g., speech recognition in 
previous studies. The covariance matrices in MVDR are estimated using time-frequency masks 
generated by two components complex Gaussian mixture models~\cite{higuchi2016robust}.
Given test audio samples $ a_{j}\in A=\{a_{1},a_{2},...a_{M}\}$, after 
performing WPE and MVDR on utterance $a_{j}$, we get enhanced 
audio set $A'=\{a'_{1},a'_{2},...a'_{M}\}$. Then 30-dimensional MFCCs 
feature is extracted from $A'$ and fed into the well-trained 
embedding extractor $F(x)$ to obtain the new embedding for verification. 
The experiments results are in Table \ref{tab:front-end}.
\begin{table}[h]
\setlength{\abovecaptionskip}{0.2cm}
\caption{Experiments results on ResNet-BAM-DAT with front-end methods on development dataset}
\label{tab:front-end}
\resizebox{\linewidth}{!}{
\begin{tabular}{cccccccc}

\hline
\multirow{2}{*}{WPE} & \multirow{2}{*}{Beamformer} & \multirow{2}{*}{DAT} & \multicolumn{1}{c|}{\multirow{2}{*}{\begin{tabular}[c]{@{}c@{}}Data\\ Selection\end{tabular}}} & \multicolumn{2}{c|}{Task1} & \multicolumn{2}{c}{Task3} \\ \cline{5-8} 
 &  &  & \multicolumn{1}{c|}{} & EER (\%) & minDCF & EER (\%) & minDCF \\ \hline
$\checkmark$ & $\times$ & $\times$ & $\times$ & 7.12 & 0.7820 & 6.83 & 0.7216 \\
$\times$ & $\checkmark$ & $\times$ & $\times$ & 7.21 & 0.8177 & 6.92 & 0.7541 \\
$\checkmark$ & $\times$ & $\times$ & $\checkmark$ & 6.46 & 0.6901 & 5.97 & 0.6728 \\
$\times$ & $\checkmark$ & $\checkmark$ & $\times$ & 6.39 & 0.6873 & 5.89 & 0.6538 \\
$\checkmark$ & $\checkmark$ & $\checkmark$ & $\checkmark$ & \textbf{6.22} & \textbf{0.6518} & \textbf{5.86} & \textbf{0.6006} \\ \hline
\end{tabular}}
\end{table}

We find that there is no improvement but worse results on the use of the 
front-end processing pipeline. Hence we double-check the experimental settings and find 
that after the enhancement processing some audios are contaminated with stronger noise due to 
the failure of the original microphone channels, which accounts for 10\% of the total 
development set. To deal with the problem, we propose a data selection strategy 
to exclude the failure utterances from the processed data. We compare the
cosine distance between original embedding and enhanced embedding and discard the enhanced 
data which score is lower than a threshold $\theta$ 
(empirically set to 0.7 in our experiments). This way can ensure the 
processed data not too outrageous and avoid performance degradation. 

\begin{figure}[h]
  \centering
  \includegraphics[width=\linewidth]{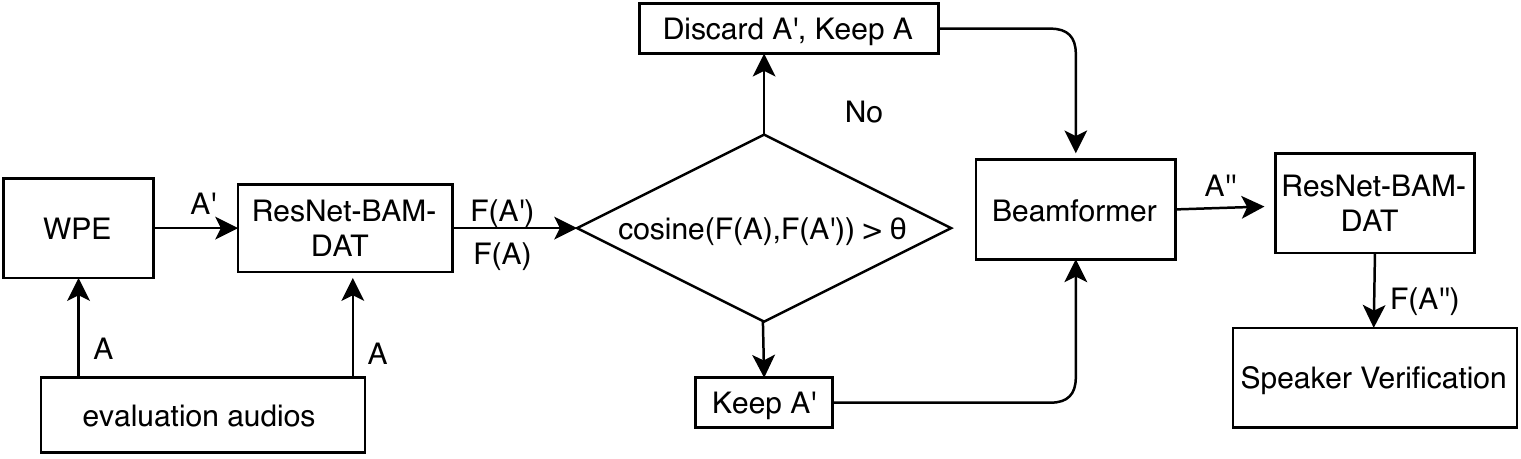}
  \caption{Evaluation flow with front-end  technologies.}
  \label{fig:front-end}
\end{figure}

Data processed by beamformer, paired with iPhone-recorded data are
used to fine-tune the ResNet-BAM-DAT model again. Note that the two domains for adversarial training are close-talk speech and far-field speech. The evaluation flow 
is shown in  Figure~\ref{fig:front-end}. After the data selection and another 
round of model fine-tuning, we achieve reasonable results on the development set, as shown in Table~\ref{tab:front-end}. With WPE and MVDR beamforming, we can achieve 0.5\% EER absolute reduction. 

\subsection{Data augmentation and selection}
Data augmentation is a commonly strategy to improve the data coverage. 
We use open-sourced MUSAN~\cite{snyder2015musan} noise and 
room impulse response (RIR) databases from \cite{allen1979image} to 
perform training data augmentation, using the official scripts provided by Kaldi. At the same time, we do voice variable speed augmentation on enrollment and test of development and evaluation data.

Besides, to match the far-field conditions, we simulate multi-channel version of the enrollment data using artificially generated RIRs. We generate totally 40,000 RIRs from 200 different room configurations and the same configurations of microphone arrays as described in \cite{qin2020ffsvc}, which aims to cover the recording environment of the challenge data. After the simulation procedure, we adopt a different selection strategy from Section 6.1 to pick up well quality simulated far-field utterances. We has a clever strategy to determine hyperparameter $\theta$ which is not a fixed empirical data but dependent on the development data. Development test trials with labels and evaluation test trials without labels~\cite{qin2020ffsvc} have recorded in the same room and
devices. Acoustic properties caused by far-field scenes are the same. We use simulated development data doing cosine distance with corresponding speaker’s test data (far-field) to obtain a series suitable parameters about generating RIR and a good hyperparameter $\theta$. We use the fit RIR parameters to simulate enrollment data of evaluation data and then use the appropriate $\theta$ to select high quality simulated data to do test. This way can find more benefit RIR parameters to fit the real recorded environment of test data to reduce the mismatch between 
simulated enrollment and test data. 

The effects of data augmentation are shown in Table~\ref{tab:data_aug_resulet}. Augmentation on training data brings 1\% reduction on EER. 
Meanwhile, doing data augmentation on enrollment and test set, EER is reduced the most, even up to 2\%.

\begin{table}[h]
\setlength{\abovecaptionskip}{0.2cm}
\caption{Experiment results with data augmentation on ResNet-BAM on development dataset
}
\label{tab:data_aug_resulet}
\resizebox{\linewidth}{!}{
\begin{tabular}{ccccccc}
\hline
\multirow{2}{*}{\begin{tabular}[c]{@{}c@{}}Aug.\\ Training\end{tabular}} & \multirow{2}{*}{\begin{tabular}[c]{@{}c@{}}Aug.\\ Enrollment\end{tabular}} & \multicolumn{1}{c|}{\multirow{2}{*}{\begin{tabular}[c]{@{}c@{}}Aug. \\ Test\end{tabular}}} & \multicolumn{2}{c|}{Task1} & \multicolumn{2}{c}{Task3} \\ \cline{4-7} 
 &  & \multicolumn{1}{c|}{} & EER (\%) & minDCF & EER (\%) & minDCF \\ \hline
$\checkmark$ & $\times$ & $\times$ & 6.37 & 0.6911 & 5.45 & 0.5622 \\
$\checkmark$ & $\checkmark$ & $\times$ & 4.81 & 0.5009 & 3.55 & 0.3977 \\
$\checkmark$ & $\checkmark$ & $\checkmark$ & \textbf{4.22} & \textbf{0.4213} & \textbf{3.39} & \textbf{0.3728} \\ \hline
\end{tabular}}
\end{table}




\section{Evaluation Results and Conclusions }
At the middle deadline, our submitted system is ResNet-BAM trained with augmented (and selected) data. The evaluation results of task 1 and task 3 on the leadboard of FFSCV Official website are shown in Table ~\ref{tab:eval_results}. Our submission achieves the first and second place on task 1 and 3 respectively. The scores of our submitted systems are bold and red-colored.

\begin{table}[h]
\setlength{\abovecaptionskip}{0.2cm}
\scriptsize
\caption{Evaluation dataset results of the top 3 teams by middle deadline on leadboard.}
\label{tab:eval_results}
 \centering
\begin{tabular}{ccccc}
\hline
\multicolumn{1}{c|}{\multirow{2}{*}{Rank}} & \multicolumn{2}{l|}{Task1} & \multicolumn{2}{c}{Task3} \\ \cline{2-5} 
\multicolumn{1}{c|}{} & EER (\%) & minDCF & EER (\%) & minDCF \\ \hline
1&\textbf{\textcolor{red}{5.39}} &\textbf{\textcolor{red}{0.4636}} & 5.53 & 0.4584 \\
2 & 5.08 & 0.5002 & \textbf{\textcolor{red}{6.44}} & \textbf{\textcolor{red}{0.4585}} \\
3 & 4.72 & 0.5200 & 5.14 & 0.4708 \\ \hline
\end{tabular}
\end{table}

This paper introduces the main approaches used in our submitted system to FFSVC, specially designed speaker embedding network ResNet-BAM, domain adversarial training, front-end processing and data augmentation as well as selection. The most profitable method is data augmentation with data selection. There is still potential space to improve in far-field SV. For instance, we expect front-end processing should play a vital role in dealing with the mismatch problem. Especially, we plan to explore the recent neural front-end approaches, such as neural dereveberation~\cite{wang2018investigating} and neural beamformers~\cite{li2016neural,wu2019frequency}. Moreover, we hope the performance of far-field SV can be further boosted through front-end and speaker embedding network joint training.

\bibliographystyle{IEEEtran}
\bibliography{mybib.bib}
\end{document}